\documentclass{kapproc} 

\usepackage[dvips]{graphicx}

\upperandlowercase

\kluwerbib

\begin{document}

\def\etal{et al. }
\def\apj{ApJ}
\def\apjl{ApJ\ Lett.}
\def\apjs{ApJ\ Suppl.}
\def\aas{A\&A Supp.}
\def\aa{A\&A}
\def\aal{A\&A Lett.}
\def\mnras{MNRAS}
\def\mnrasl{MNRAS Lett.}
\def\inpress{in press}
\def\inprep{in prep.}
\def\submit{submitted}
\def\mjysr{MJy/sr }
\def\inu{{I_{\nu}}}
\def\fnu{{F_{\nu}}}
\def\bnu{{B_{\nu}}}
\def\mic{$\mu$m}

\articletitle[The Galactic dust as a foreground to Cosmic Microwave Background maps]
{The Galactic dust as a foreground to Cosmic Microwave Background maps}

\author{X. Dupac (ESA-ESTEC, Noordwijk, the Netherlands, xdupac@rssd.esa.int), J.-P. Bernard, N. Boudet, M. Giard (CESR Toulouse), J.-M. Lamarre (LERMA Paris), C. M\'eny (CESR), F. Pajot (IAS Orsay), I. Ristorcelli (CESR)}

\anxx{Dupac, X., et al.}

\begin{abstract}
We present results obtained with the PRONAOS balloon-borne experiment
on interstellar dust.
In particular, the submillimeter / millimeter spectral index is
found to vary between roughly 1 and 2.5 on small scales (3.5$'$ resolution). This could
have implications for component separation in Cosmic Microwave Background
maps.
\end{abstract}


\vspace*{1em}

\section{Introduction}

To accurately characterize dust emissivity properties represents a major challenge of nowadays astronomy.
It is crucial for deriving very accurate maps of the Cosmic Microwave Background fluctuations, as well as for understanding the physics of the interstellar medium.
In the submillimeter domain, large grains at thermal equilibrium (e.g. \cite{desert90}) dominate the dust emission.
This thermal dust is characterized by a temperature and a spectral dependence of the emissivity which is usually simply modelled by a spectral index.
The temperature, density and opacity of a molecular cloud are key parameters which control the structure and evolution of the clumps, and therefore, star formation.
The spectral index ($\beta$) of a given dust grain population is directly linked to the internal physical mechanisms and the chemical nature of the grains.

It is generally admitted from Kramers-K\"onig relations that 1 is a lower limit for the spectral index.
$\beta$ = 2 is particularly invoked for isotropic crystalline grains, amorphous silicates or graphitic grains.
However, it is not the case for amorphous carbon, which is thought to have a spectral index equal to 1.
Spectral indices above 2 may exist, according to several laboratory measurements on grain analogs.
Observations of the diffuse interstellar medium at large scales favour $\beta$ around 2 (e.g. \cite{boulanger96}, \cite{dunne01}).
In the case of molecular clouds, spectral indices are usually found to be between 1.5 and 2.
However, low values (0.2-1.4) of the spectral index have been observed in circumstellar environments, as well as in molecular cloud cores.

\section{PRONAOS observations\label{obs}}

PRONAOS (PROgramme NAtional d'Observations Submillim\'etriques) is a French
balloon-borne submillimeter experiment (\cite{ristorcelli98}).
Its effective wavelengths are 200, 260, 360 and 580 \mic, and the angular resolutions are 2$'$ in bands 1 and 2, 2.5$'$ in band 3 and 3.5$'$ in band 4.
The data analyzed here were obtained during the second flight of PRONAOS in September 1996, at Fort Sumner, New Mexico.
The data processing method, including deconvolution from chopped data, is described in Dupac \etal (2001).
This experiment has observed various phases of the interstellar medium, from diffuse clouds in Polaris (\cite{bernard99}) and Taurus (\cite{stepnik03}) to massive star-forming regions in Orion (\cite{ristorcelli98}, \cite{dupac01}), Messier 17 (\cite{dupac02}), Cygnus B, and the dusty envelope surrounding the young massive star GH2O 092.67+03.07 in NCS.
The $\rho$ Ophiuchi low-mass star-forming region has also been observed, as well as the edge-on spiral galaxy NGC 891 (\cite{dupac03b}).

\section{Analysis of the Galactic dust emission}

We fit a modified black body law to the spectra: $\inu = \epsilon_0 \; \bnu(\lambda,T) \; (\lambda/\lambda_0)^{-\beta}$, where $\inu$ is the spectral intensity (MJy/sr), $\epsilon_0$ is the emissivity at $\lambda_0$ of the observed dust column density, $\bnu$ is the Planck function, $T$ is the temperature and $\beta$ is the spectral index.
In most of the areas, we use either only PRONAOS data or PRONAOS + IRAS 100 \mic~data.
We restrain the analysis to all fully independent (3.5$'$ side) pixels for which both relative errors on the temperature and the spectral index are less than 20\%.
This procedure allows to reduce the degeneracy effect between the temperature and the spectral index.
Dupac \etal (2001) and Dupac \etal (2002) have shown by fitting simulated data that this artificial anticorrelation effect was small compared to the effect observed in the data.

\begin{figure}[!ht]
\includegraphics[scale=.6]{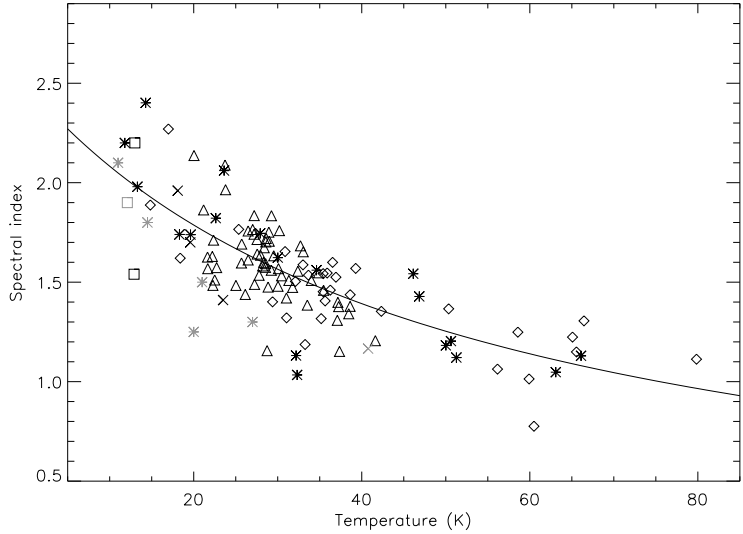}
\caption[]{Spectral index versus temperature, for fully independent pixels in Orion (black asterisks), M17 (diamonds), Cygnus (triangles), $\rho$ Ophiuchi (grey asterisks), Polaris (black squares), Taurus (grey square), NCS (grey cross) and NGC 891 (black crosses).
The full line is the result of the best hyperbolic fit: $\beta = {1 \over 0.4 + 0.008 T}$
}
\label{compil}
\end{figure}

We present in Fig. \ref{compil} the spectral index - temperature relation observed.
The temperature in this data set ranges from 11 to 80 K, and
the spectral index also exhibits large variations from 0.8 to 2.4.
One can observe an anticorrelation on these plots between the temperature and the spectral index, in the sense that the cold regions have high spectral indices around 2, and warmer regions have spectral indices below 1.5.
In particular, no data points with $T >$ 35 K and $\beta >$ 1.6 can be found, nor points with $T <$ 20 K and  $\beta <$ 1.5.
This anticorrelation effect is present for all objects in which we observe a large range of temperatures, namely Orion, M17, Cygnus and $\rho$ Ophiuchi.
It is also remarkable that the few points from other regions are well compatible with this general anticorrelation trend.
The temperature dependence of the emissivity spectral index is well fitted by a hyperbolic approximating function.

Several interpretations are possible for this effect: one is that the grain sizes change in dense environments, another is that the chemical composition of the grains is not the same in different environments and that this correlates to the temperature, a third one is that there is an intrinsic dependence of the spectral index on the temperature, due to quantum processes such as two-level tunneling effects.
Additional modeling, as well as additional laboratory measurements and astrophysical observations, are required in order to discriminate between these different interpretations.
More details about this analysis and the possible interpretations can be found in Dupac et al. (2003a).

\section{Implications for high-redshift far-infrared observations}

A recent paper from Eales et al. (2003) showed that the 850 / 1200 \mic~ratio of their sample of extragalactic millimeter sources was very low, which could be explained by spectral indices of the dust around 1 in high-redshift galaxies.
Though this result is still uncertain, it might confirm our measurements because high-redshift galaxies are likely to be warmer than low-redshift galaxies (because of cosmic expansion).

The anticorrelation between the temperature and the emissivity spectral index can indeed have major implications for deriving the redshifts, masses, temperatures, luminosities, etc, of extragalactic objects.

\section{Implications for Cosmic Microwave Background measurements}

Even high Galactic latitude clouds can have a non-uniform spectral index (e.g. \cite{bernard99}).
Since the Galactic dust is the major contributor to CMB foregrounds in the submillimeter and millimeter domains, even small variations of the dust spectral energy distribution can harm the CMB measurements if the component-separation methods assume a uniform dust spectral index.

\end{document}